\documentstyle[12pt]{article}
\newcommand{\cftnote}
{\renewcommand{\thefootnote}{\fnsymbol{footnote}}}
\newcommand{\resetftnote}{\setcounter{footnote}{0}}

\begin{document}

\begin{flushright}
UVA-93-41\\
December 1993\\
%{\large DRAFT}
\end{flushright}

\vspace{2mm}

\begin{center}
\cftnote

{\Large \bf SINGULARITY THEORY FOR W-ALGEBRA POTENTIALS}
\footnote{Research supported in part by CICYT
}\\[7mm]
{\large Jos\'e Gaite}
\\[4mm]
{\it Instituut voor Theoretische Fysica, University of Amsterdam,\\
Valckenierstraat 65, 1018 XE Amsterdam, The Netherlands}\\[2mm] and \\[2mm]
{\it Instituto de Matem\'aticas y
F\'{\i}sica Fundamental, C.S.I.C.,\\ Serrano 123, 28006 MADRID,
Spain.}\\
\resetftnote
\end{center}

\vspace{2mm}

\begin{abstract}
The Landau potentials of $W_3$-algebra models are analyzed with
algebraic-geometric methods. The number of ground states and the number of
independent perturbations of every potential coincide and can be
computed. This number agrees with the structure of ground states
obtained in a previous paper, namely, as the phase structure of
the IRF models of Jimbo et al.
The singularities associated to these potentials are identified.
\end{abstract}

\global\parskip 6pt

\section{Introduction}

After the remarkable work of Zamolodchikov
(see also Ludwig-Cardy) \cite{ZamPot}
showing how to construct Landau-Ginzburg potentials for the minimal
models of 2d CFT, there have been many interesting extensions and
generalizations. In particular, we saw in a previous paper that
Zamolodchikov's method with some qualifications works perfectly for
the natural generalization
of the Virasoro minimal models to higher discrete symmetry, namely, the
minimal models of W-algebras.
We found the corresponding lagrangians and analyzed the phase structure
that they entail. The chief result was the construction of a perturbation
that produces the desired state diagrams, namely, dominant-weight diagrams
of the corresponding Lie algebra, thus reproducing the phase structure
of the IRF models of Jimbo et al \cite{JiMi}.
In the case of $W_{(3)}$, they are
triangular diagrams with a particular structure~\cite{NPB}.

However, in absence of further analysis we cannot assert that the
state diagram mentioned above is the maximum possible unfolding
of ground states. Neither do we know that the algebra of perturbations
corresponds to the relevant fields of a definite model. In fact,
for all we know, it might even not be finite. In comparison,
these problems do not arise for the potentials of the minimal
Virasoro models, since they coincide with the ADE singularities,
the resolution of which is standard in the literature \cite{Arnold}.

This paper is devoted to establish the unfolding of extremal points and
dimension of perturbation algebras of $W_{(3)}$-potentials and
to prove therefore that they are well-defined and just right to
describe the desired phase structure. We should point out that
both features, namely, number of extremal points and dimension
of the perturbation algebra, are intimately related and actually coincide,
as will be explained below. We will work out in detail the low-$p$ cases,
explicitly exhibiting the perturbation algebras.
Finally, we find that these potentials correspond to
a simple series of two-variable singularities, of which the first two are
already known.

\section {Ground states and algebra of perturbations. General methods}

Let us recall the Landau potential for $W_{(3)}^p$
\begin{eqnarray}
V &\simeq&
\sigma^{p-4}\, \bar{\sigma}^{p-1}
+ (\sigma \bar{\sigma})^{p-2} + c.c.  \nonumber\\
&=&
(\sigma \bar{\sigma})^{p-4}\, \left(\sigma^3 +\bar{\sigma}^3\right) +
(\sigma \bar{\sigma})^{p-2} .            \label{oldpotential}
\end{eqnarray}
{}From now on, we consider the complexified potential, letting
${\sigma}$ and $\bar{\sigma}$ take arbitrary complex values and
we rename them as x and y, respectively\footnote
{Do not confuse with the usual notation in which x and y are
{\em real} variables
defined by ${\sigma}=x+iy$, as in \cite{NPB}.}.
Since numerical coefficients were disregarded in
obtaining~(\ref{oldpotential})
and they not play any role in this paper either, we will choose them
as to simplify the calculations.
Thus we write the potential as
\begin{equation}
V = {{{{\left( x\,y \right) }^{p -2 }}}\over {p -2}} -
{{\left( x\,y \right) }^{p -4 }}\,\left( {x^3} + {y^3} \right).
\label{potential}
\end{equation}
It gives two algebraic equations for the equilibrium
(extremal) points,
\begin{eqnarray}
V_x \equiv \partial_x V =  y\,{{\left( x\,y \right) }^{-5 + p}}\,
   \left(\left( -p+1 \right) \,{x^3} + {x^2}\,{y^2} +
     \left( -p+4 \right) \,{y^3} \right) = 0,    \label{eqx}\\
V_y \equiv \partial_y V = x\,{{\left( x\,y \right) }^{-5 + p}}\,
   \left( \left( -p+4 \right) \,{x^3} + {x^2}\,{y^2} +
     \left( -p+1\right) \,{y^3} \right) = 0.   \label{eqy}
\end{eqnarray}
Of course, we are only interested in the real solutions, such that
$y = {\bar x}$ and the equations are conjugate of one another.
Since they are of degree $2p-5$, one should expect either a maximum of
$(2p-5)^2$ solutions or an infinite number, according to Bezout's
theorem. In geometrical language, if the two algebraic curves intersect
at a number of points larger than the product of their degrees, they
must have a branch in common. This latter possibility must be excluded;
it implies that the potential is not determinate, namely, there are
higher degree terms that cannot be removed by diffeomorphisms \cite{Catas}.
However, it is clearly the general case in~(\ref{eqx}, \ref{eqy}),
where the common factor $x\,y$ appears in both equations when $p \geq 6$.
This problem was already anticipated in \cite{NPB}, as well as its
solution, namely, to consider further terms in the potential.
We will come back to it in section 5,
where we construct the remaining terms.

On the other hand, the number $(2p-5)^2$ of solutions is still larger
than the number of extremal points in the state diagram we seek to
reproduce, as will be seen more concretely below.
Nevertheless, it can still happen that the structure of those equations
is such that the maximum number of solutions is never reached.
To understand why, it is convenient to resort to the geometrical picture:
Although the two algebraic curves must have the $(2p-5)^2$ intersection
points (counting multiplicities) in projective space, it can occur that
some of them are permanently placed at infinity. Let us see that it is
the case. If we homogenize the equations by adding a third variable $z$
and we make $z=0$ to find the points at infinity, we inmediatly see
that some of them are common to both curves. Furthermore, this property
is not altered when we introduce perturbations.

The method we will use to establish the actual number of extremal
points is half way to solve Eqs.~(\ref{eqx}, \ref{eqy}).
The way one should attempt to
solve a pair of equations with two unknown variables is by first
eliminating one variable and, therefore, one equation, thus reducing
the problem to an ordinary equation in one variable.
The systematic way to perform this operation utilizes an
algebraic object called resultant.
The resultant of two polynomials in one variable is defined as
the simplest algebraic object that vanishes whenever they have
a common root; namely, the product of differences of roots
of each polynomial. Since it is symmetric in each set of roots,
it can be expressed in terms of the coefficients of the polynomials.
This expression is obtained in algebra textbooks but will not be needed.
Now, if we regard our two polynomials~(\ref{eqx}, \ref{eqy}) as
polynomials in one variable, y say, with coefficients that are
polynomials in the other, x, their resultant is also a polynomial in x.
Moreover, for those values of x that make it null
the Eqs.~(\ref{eqx}, \ref{eqy})
will be satisfied. In fact, according to our purpose,
we shall not attempt to solve this equation in x but rather to
determine the number of solutions; this can be done just by inspection.

We consider now the algebra of perturbations of a potential $V$.
This algebra is given by the quotient of all possible deformations
by those that are induced by diffeomorphisms \cite{Catas},
\[Q = {P[x,y]\over{(V_x,V_y)}}.\]
Alternatively, we can consider, in place of the potential, the
curves~(\ref{eqx},\ref{eqy}).
Each curve has naturally associated to it the ideal
of polynomial functions that vanish on it, namely $(V_x)$ or $(V_y)$.
The algebras of polynomial functions on the curves are
$Q_x = P[x,y]/(V_x)$ and
$Q_y = P[x,y]/(V_y)$, respectively.
The algebra of polynomials on the intersection is $Q$.
Since we are assuming that this intersection is just a finite set of
points, $Q$ is finite and contains as many elements as there are
intersection points (counting multiplicities) \cite{Fulton}.

In conclusion, we have a suitable method to obtain
the number of extremal points or elements of the perturbation algebra.
Nevertheless, to actually determine the ones or the others, one is to
undertake detailed computations: Respectively, to completely solve
the equations~(\ref{eqx}, \ref{eqy})
to find the extremal points or to use these equations
to eliminate all perturbations
that they generate and see which ones are left.

\section{The $W_{(3)}^4$ potential and its relation to the $D_4$
singularity}

The potential for the first model in the series, $W_{(3)}^4$, is well known.
It was described in \cite{JPA} in the context of symmetric catastrophe
theory. There were found the six relevant perturbations; with the
addition of the identity, a total number of seven independent perturbations.
Correspondingly, it was shown that the unfolding consists of seven points
in a triangular array that constitutes the first instance of the series
later analyzed in \cite{NPB}. We now intend to reobtain
these results within our present notation and objectives.

The potential
\begin{equation}
V = {{{x^2}\,{y^2}}\over 2} - {{{x^3} + {y^3}}\over 3} \\
\end{equation}
produces the equations
\begin{eqnarray}
V_x =  -{x^2} + x\,{y^2} = 0  \label{eqx.4}\\
V_y =  {x^2}\,y - {y^2} = 0     \label{eqy.4}
\end{eqnarray}
Given their simplicity, we can actually solve them.
We thus have (0,0) with multiplicity 4 and 3 other,
$(x=$ cubic root of 1, $y=x^2)$, seven solutions altogether.
It is indeed fewer than
the expected number for two third-degree equations, which is nine.
A more convenient way to realize this fact is to compute the
resultant\footnote
{We use the Mathematica object Resultant.},
\begin{equation}
{\rm Resultant}[V_x,V_y, y] = {x^4} - {x^7}.
\end{equation}
{}From the geometrical point of view,
the cause is clear: The curves~(\ref{eqx.4}, \ref{eqy.4})
comprise two parabolas and two
intersection points are placed at infinity.

In this case, it is also straightforward to calculate the perturbation
algebra $Q$. It is clear that
it contains all monomials up to the second degree,
$1, x, y, x^2, y^2, x\,y$. At the third degree, we have to exclude
right away two monomials, $x\,{y^2}, {x^2}\,y$,
because of Eq.~(\ref{eqx.4}, \ref{eqy.4}).
Furthermore, we can easily obtain ${{x^3} = {y^3}}$, thus only leaving
${{x^3} + {y^3}}$. Now, it is not difficult to see that every remaining
monomial can be expressed in terms of the ones just selected.
For example, $x^4 = x\,y^3 = x^2\,y =y^2$.

We must remark that, strictly speaking, $x^3+y^3$ cannot be considered
a perturbation for if it is removed from the potential, the
remaining quartic term is the degenerate double-cusp catastrophe,
an indeterminate potential. On the other hand, the quartic term itself
can be safely removed: the cubic term constitutes by itself a
well defined and known potential, the $D_4$ singularity of the ADE
classification. We can further appreciate this point by observing
that the unperturbed potential has a quadruple extremal point at
the origin and three distinct others. The latter are obviously
irrelevant for the singularity, although not for the potential
as a whole, since they are precisely the ground states.

\section{The $W_{(3)}^5$ potential. Arnold's $N_{16}$ singularity}

The next case, $W_{(3)}^5$, has the potential
\begin{equation}
V = {{{x^3}\,{y^3}}\over 3} - x\,y\,\left( {x^3} + {y^3} \right)
\end{equation}
which has not been thorouhgly analyzed before. We know, however, that
a definite symmetric perturbation will produce the desired triangular
array of 19 extremal points, of which 6 are minima \cite{NPB}.
Proceeding as above,
we consider the equations
\begin{eqnarray}
V_x = -4\,{x^3}\,y + {x^2}\,{y^3} - {y^4} = 0,    \label{eqx.5} \\
V_y = -{x^4} + {x^3}\,{y^2} - 4\,x\,{y^3} = 0.\label{eqy.5}
\end{eqnarray}
Despite their unfriendly aspect, they are solvable. However, we must
not bother to actually solve them, since we can get similar information
from the resultant,
\begin{equation}
{\rm Resultant}[V_x,V_y, y] = 3375\,{x^{16}} - 27\,{x^{19}} \label{resul.5}
\end{equation}
We see that we have indeed a maximum of 19 solutions, 16 of which
are degenerate at (0,0), while the other 3 are again cubic roots.

We are now to calculate the perturbation algebra $Q$.
Right away, we can count in
the 15 monomials up to the fourth degree. At the fifth degree, we must
exclude two, according to~(\ref{eqx.5}, \ref{eqy.5}).
Furthermore, less obviously,
$x^4\,y = y^4\,x$. Thus we have left $x^5, y^5, x^4\,y + y^4\,x$.
At sixth degree, there is an independent one yet, $x^6 + y^6$, whereas
the remaining and any of higher degree are reducible to those already
mentioned. The total number is 19, of course.

We have to face now the problem analogous
to that of the previous potential;
namely, we cannot consider $x^4\,y + y^4\,x$ as a perturbation, for
the same reason. Moreover, we have three other terms,
$x^5, y^5$, and $x^6+y^6$, that do not fit the expected perturbations
for this potential. We are thus well advised to resort to singularity
theory to further analyze this potential. In analogy with the previous
case, we may expect the term $x^4\,y + y^4\,x$ to furnish the
appropriate singularity. Let us see whether it has the correct
properties. First of all, we know from~(\ref{resul.5}) that its
critical point must be 16-fold degenerate. It is easy to convince
oneself that removing the 6th degree term from the potential
implies reducing the resultant to just $x^{16}$. Besides, the candidate
singularity has already been identified and named $N_{16}$;
it has codimension 12 and 3 modular parameters, associated to
the monomials $x^2\,y^3, y^2\,x^3$ and $x^3\,y^3$ \cite{Arnold}.

Therefore, we now realize that the term $x^3\,y^3$ in the potential is
just a modular deformation that needs not be present\footnote
{We must remark again though that it is essential for the
potential to have absolute minima and, in particular, the minima
that we expect.}.
Then, the equations~(\ref{eqx.5}, \ref{eqy.5})
become homogeneous of 4th degree,
and indicate that two monomials of this degree do not belong to
the perturbation algebra; let them be $x^4$ and $y^4$. The remaining 13
monomials up to degree 4th account for the codimension 12. Besides,
there are two other monomials in the perturbation algebra,
$x^2\,y^3 = x^5/4$ and $y^2\,x^3 = y^5/4$, corresponding
to modular deformations.

It is pertinent to point out that there is another form of the $N_{16}$
singularity more suited to computations, namely $x^5 + y^5$. However,
it does not have $D_3$ symmetry (it has $D_5$ instead) and is thus not
convenient for our purposes.

\section{$W_{(3)}^6$ and the general $W_{(3)}^p$ case. The series $W^l$ of
singularities}

The potential of $W_{(3)}^6$,
\begin{equation}
V = {x^4}\,{y^4} - {x^2}\,{y^2}\,\left( {x^3} + {y^3} \right) \label{pot.6}
\end{equation}
is incomplete, as was mentioned in section 2. It is easy to see
that essentially we have only two possible additions with
the necessary symmetry,
${{{x^6} + {y^6}}}$
or $x\,y\,{\left({x^6} + {y^6}\right)}$.
The second one yields a resultant
of degree 49th,
showing that the number of extremal points is the maximum allowed by
the degree of the potential. The first one yields
\begin{equation}
{\rm Resultant}[V_x,V_y, y] =
-1024\,{x^{25}} + 23168\,{x^{28}} - 25120\,{x^{31}}
+ 8944\,{x^{34}} -  1024\,{x^{37}},   \label{resul.6}
\end{equation}
with 37 extrema. They account for the 10 minima, 18 saddle points and
9 maxima of the corresponding triangular array, as was depicted in
\cite[Fig. 2]{NPB}. With some effort one could equally obtain the
37 elements of $Q$ but it is not worthwile.

Let us now consider the type of singularity of this potential.
The multiplicity at the origin is 25, according to~(\ref{resul.6}).
This is precisely the multiplicity of $V={{x^6} + {y^6}}$, as can be
seen at once from the equations $V_x=0$ and $V_y=0$ \footnote
{More generally, the multiplicity of ${{x^n} + {y^n}}$ is $(n-1)^2$.}.
As far as we know,
this singularity has never been studied in the literature.
It has codimension 18 and 6 modular parameters. One of these
modular parameters corresponds to ${x^4}\,{y^4}$, of course.

We should attempt to generalize the previous results to arbitrary $p$.
The first problem is to find out the correct way to complete
the potential. To this end, we can appeal to the results in \cite{NPB}.
Let us recall that we obtained there the algebraic curve
that considered as a level curve of some potential gives the necessary
extremal-point structure. That curve was the product
\begin{equation}
\prod_{i=1}^{(l+1)/2} (I_0^2 + I_1 + w_i I_0) =
I_0^{l+1} + I_0^{l-1} I_1 + \cdots = 0,     \label{odd}
\end{equation}
for $l$ odd, or
\begin{equation}
\prod_{i=1}^{l/2} (I_0^2 + I_1 + w_i I_0)(I_0-c) =
I_0^{l+1} + I_0^{l-1} I_1 + \cdots = 0,     \label{even}
\end{equation}
for $l$ even.
The point was that the two first terms coincide with those of
the potential~(\ref{potential}) if $l=p-3$. However, it must be noticed
that there are further terms that should be regarded as belonging
to the unperturbed potential, since they remain when all
perturbation parameters, $w_i$ and $c$, are made null.
Nevertheless, there is an excessive number of
monomials in $x$ and $y$; e.g., the lowest degree term in~(\ref{odd}),
$I_1^{(l+1)/2}$, or in~(\ref{even}), $I_0\,I_1^{l/2}$, produce
when expanded a string of monomials with only two essential ones,
$x^{3(l+1)/2}+y^{3(l+1)/2}$ or $(x\,y)\,\left(x^{3l/2}+y^{3l/2}\right)$,
respectively.
Keeping only the essential terms, we are thus led to propose a potential
\begin{equation}
V = {x^{p-2}}\,{y^{p-2}}
  - {x^{p-4}}\,{y^{p-4}}\,\left( {x^3} + {y^3} \right)  +
   {x^{p-6}}\,{y^{p-6}}\,\left( {x^6} + {y^6} \right)  + \cdots +
 \left( {x^{3\,{p-2\over2}}} + {y^{3\,{p-2\over2}}} \right)\label{c.pot.e}
\end{equation}
for $p$ even or
\begin{equation}
V = {x^{p-2}}\,{y^{p-2}}
- {x^{p-4}}\,{y^{p-4}}\,\left( {x^3} + {y^3} \right)  +
   {x^{p-6}}\,{y^{p-6}}\,\left( {x^6} + {y^6} \right)  + \cdots +
x\,y\,\left( {x^{3\,{p-3\over2}}} + {y^{3\,{p-3\over2}}} \right)
\label{c.pot.o}
\end{equation}
for $p$ odd,
which nicely generalizes that of~(\ref{pot.6}) to larger p.

It remains to be proved that
this potential~(\ref{c.pot.e} or \ref{c.pot.o})
has the properties that we found for those up to $p=6$; namely,
the maximum number of extrema is as the corresponding triangular array
and the independent perturbations can be obtained accordingly.
Unfortunately, the form of~(\ref{c.pot.e} or \ref{c.pot.o})
as a sum of a number of terms proportional to $p$
makes it unsuitable for
the algebraic computation of the resultant. Nevertheless, we have performed
the calculations up to $p=11$ and checked that
the resultant has the expected degree.
The results for $p=7$ to 9 are shown in the
appendix. We believe that these partial results, beside their own interest,
provide some evidence for
the general case.

An important result can be directly read off
from~(\ref{c.pot.e}, \ref{c.pot.o}),
namely, the series of singularities associated to our potentials:
\[x^{3(l+1)/2}+y^{3(l+1)/2},~~l=1,3,5, \cdots,\] or
\[(x\,y)\,\left(x^{3l/2}+y^{3l/2}\right),~~l=2,4,6, \cdots.\]
We think that $W^l$ is a convenient name for this series\footnote
{We follow Arnold in giving names to series of singularities \cite{Arnold}.
In fact, he has already used most letters to this effect, including $W$.
Nevertheless, we think that it is the adequate letter here and that no
confusion should arise since he places the other letter(s) as subscript
whereas we have placed $l$ as superscript.};
then $W^1 = D_4$ and $W^2 = N_{16}$.
The latter form above, when $l$ is even, is equivalent to
$\left(x^{(3l/2)+2}+y^{(3l/2)+2}\right)$.
Hence, we see that both belong to
a simple series of two-variable singularities, $x^n+y^n$.
They have multiplicity $\mu = (n-1)^2$, moduli $m = (n-3)(n-2)/2$ and
codimension $c= (n+3)(n-2)/2$ such that they fulfill $\mu = c + m + 1$
\cite{Arnold}. Among these large moduli are the other terms in
the complete potential~(\ref{c.pot.e}, \ref{c.pot.o}), as well as those
inessential terms in~(\ref{odd}, \ref{even}) that were discarded.

\section{Conclusions}

The Landau potentials of $W_3$ models are analyzed in order to
ascertain whether they fit the phase structure of the IRF models
of Jimbo et al. Their sufficiency to this purpose has been
established in a previous paper \cite{NPB} by constructing a perturbation
that produces the required configuration of extrema.
This configuration is now shown to be the maximum possible
unfolding. The method used only allows to consider a finite number of cases.
The reason is that it is based on an algebraic study of the equations
for extrema, which final step is to obtain the resultant
with Mathematica. This is done sequentially for increasing $p$.
It has been carried out up to $p= 11$ but only the cases up to
$p=9$ are displayed.
The potentials for $p= 4$ and 5 deserve a more detailed study;
their algebras of perturbations are also obtained.

A necessary step to attain these results is to complete the potentials,
which are otherwise indeterminate. This is done in general, yielding
a sum of terms of decreasing degree, the last of which depends on
the parity of $p$~(\ref{c.pot.e}, \ref{c.pot.o}).
Precisely these last terms
give the singularities of these potentials, while the others can
be regarded as modular terms. Of course, this is the mathematical
viewpoint, which disregard the physical significance of the
various terms. An analysis along a more physical line should require us to
resort to the actual fields associated to the monomials according
to the identifications realized in~\cite{NPB} and is beyond the scope
of the present paper.

\vspace{8mm}
\noindent
{\large \bf Acknowledgement}\\[2mm]
I am grateful to Jean-Bernard Zuber for asking the right questions and
for reading this manuscript.

\pagebreak[2]

\appendix

\section{Appendix}
We show here that the resultants for the $p=7, 8$ and 9 potentials
precisely account for the number of extrema in the corresponding
diagrams. Let us first calculate the number of extrema in the $p$th
triangular diagram, which can be done in full generality by clever
use of triangular numbers. This diagram has $p-2$ minima on each side.
Hence, the total number of minima is $(p-2)(p-1)/2$. There is exactly
one saddle point between each pair of minima, so that saddle points and
minima form together a triangular array of side $2p-5$. Therefore, their
total number is $(2p-5)(2p-4)/2$. The distribution of maxima is a little
more involved; however, they can be split in two triangles of sides
$p-3$ and $p-4$, the total number being $(p-3)^2$. Adding both numbers,
we obtain $3 p (p-5) + 19$ extrema altogether. In particular, 61, 91
and 127 for $p=7, 8$ and 9, respectively.

Now we show the results of a computation of the resultants with
Mathematica. The potential, equations and resultant are written
for each case separately:
\begin{itemize}
\item $p=7$
\begin{equation}
V = x\,y\,\left( {x^4}\,{y^4} - {x^2}\,{y^2}\,\left( {x^3} + {y^3} \right)
     + {{{x^6} + {y^6}}\over 3} \right)
\end{equation}
\vspace{1mm}
\begin{equation}
V_x = {{7\,{x^6}\,y}\over 3} - 6\,{x^5}\,{y^3} + 5\,{x^4}\,{y^5} -
   3\,{x^2}\,{y^6} + {{{y^7}}\over 3}  = 0
\end{equation}
\begin{equation}
V_y = {{{x^7}}\over 3} - 3\,{x^6}\,{y^2} + 5\,{x^5}\,{y^4}
   - 6\,{x^3}\,{y^5} + {{7\,x\,{y^6}}\over 3} = 0
\end{equation}
\vspace{1mm}
\protect\begin{eqnarray}
{\rm Resultant}[V_x,V_y, y] = {12230590464\over {1594323}}\,{x^{49}} -
   {246594502656\over {1594323}}\,{x^{52}}  \nonumber \\
      + {249221422080\over {1594323}}\,{x^{55}}
       - {81199756125\over {1594323}}\,{x^{58}} +
     {8303765625\over {1594323}}\,{x^{61}}
\protect\end{eqnarray}
\item $p=8$
\begin{equation}
V = {x^6}\,{y^6} - {x^4}\,{y^4}\,\left( {x^3} + {y^3} \right)  +
   {x^2}\,{y^2}\,\left( {x^6} + {y^6} \right)  + {{{x^9} + {y^9}}\over 3}
\end{equation}
\vspace{1mm}
\begin{equation}
V_x = 3\,{x^8} + 8\,{x^7}\,{y^2} - 7\,{x^6}\,{y^4} + 6\,{x^5}\,{y^6} -
   4\,{x^3}\,{y^7} + 2\,x\,{y^8} = 0
\end{equation}
\begin{equation}
V_y = 2\,{x^8}\,y - 4\,{x^7}\,{y^3} + 6\,{x^6}\,{y^5} - 7\,{x^4}\,{y^6} +
   8\,{x^2}\,{y^7} + 3\,{y^8}  = 0
\end{equation}
\vspace{1mm}
\begin{eqnarray}
{\rm Resultant}[V_x,V_y, y] = 43046721\,{x^{64}} + 2027978856\,{x^{67}}-
   75467102058\,{x^{70}} \nonumber \\  +
   2349506627136\,{x^{73}} - 26268032446551\,{x^{76}}
   +172417135016664\,{x^{79}} \nonumber \\  + 42570239554944\,{x^{82}} +
   11049268755456\,{x^{85}} +
 577322090496\,{x^{88}} \nonumber \\ + 990677827584\,{x^{91}}
\end{eqnarray}
\item $p=9$
\begin{equation}
V =  {x^7}\,{y^7} - {x^5}\,{y^5}\,\left( {x^3} + {y^3} \right)  +
   {x^3}\,{y^3}\,\left( {x^6} + {y^6} \right)  +
   x\,y\,\left( {x^9} + {y^9} \right)  \\
\end{equation}
\vspace{1mm}
\begin{equation}
V_x = 10\,{x^9}\,y + 9\,{x^8}\,{y^3} - 8\,{x^7}\,{y^5} + 7\,{x^6}\,{y^7}
   - 5\,{x^4}\,{y^8} + 3\,{x^2}\,{y^9} + {y^{10}} = 0
\end{equation}
\begin{equation}
V_y = {x^{10}} + 3\,{x^9}\,{y^2} - 5\,{x^8}\,{y^4} + 7\,{x^7}\,{y^6} -
   8\,{x^5}\,{y^7} + 9\,{x^3}\,{y^8} + 10\,x\,{y^9} = 0
\end{equation}
\vspace{1mm}
\begin{eqnarray}
{\rm Resultant}[V_x,V_y, y] = -913517247483640899\,{x^{100}}
        - 2700744078773949615\,{x^{103}}  \nonumber            \\
 + 14078078713498713606\,{x^{106}}
- 54656571881643296139\,{x^{109}}\nonumber \\+
62947451766811208022\,{x^{112}}
- 62526864228499636560\,{x^{115}} \nonumber \\-
8771826922328936340\,{x^{118}}
- 1233504732780966240\,{x^{121}} \nonumber \\-
 38877206300966880\,{x^{124}} - 67659951487955904\,{x^{127}}
\end{eqnarray}
\end{itemize}


\begin{thebibliography}{99}

\bibitem{ZamPot} A.B. Zamolodchikov, Sov. J. Nucl. Phys. 44 (1987)
529\\
A.W.W. Ludwig and J.L. Cardy, Nucl. Phys B285 [FS19] (1987) 687

\bibitem{JiMi} M. Jimbo, T. Miwa and M. Okado, Nucl. Phys B300
(1988) 74

\bibitem{NPB} J. Gaite,
{\em Landau-Ginzburg lagrangians for W-algebra models},
UTTG-15-92 report, hep-th/9208003, to appear
in Nucl. Phys. B 411 (1994) 321

%\bibitem{KoYa} I.G. Koh and S.K. Yang, Phys. Lett. B223 (1989) 349

%\bibitem{N = 2} E. Martinec, {\it Criticality, catastrophes and
%compactifications}, in {\it Physics and mathematics of strings}, V.G.
%Knizhnik memorial volume, L. Brink, D. Friedan and A.M. Polyakov eds.,
%World Scientific 1990;\\
%N. Warner, $N = 2$ {\it Supersymmetric Integrable Models and Topological
%Field Theories}, to appear in the proceedings of the 1992 Trieste Summer
%School

\bibitem{Arnold} V.I. Arnold, S.M. Gusein-Zade and A.N. Varchenko,
{\em Singularities of Differentiable Maps}, Vol. 1 (Birkh\"auser,
Basel, 1985)

\bibitem{Catas} T. Poston and I.N. Stewart, {\em Catastrophe Theory
and Its Applications} (Pitman, London, 1976)

\bibitem{Fulton} W.F. Fulton, {\em Algebraic Curves} (W.A. Benjamin,
New York, 1969)

\bibitem{JPA} J. Gaite, J. Phys. A25 (1992) 3051

\end{thebibliography}
\end{document}